\documentclass[12pt,twoside]{article}
\usepackage{fleqn,espcrc1}

\usepackage{graphicx}
\usepackage{epsfig}
\usepackage[figuresright]{rotating}

\newcommand{\AmS}{{\protect\the\textfont2
  A\kern-.1667em\lower.5ex\hbox{M}\kern-.125emS}}

\title{Feynman-Schwinger Representation method for bound states}

\author{\c{C}etin \c{S}avkl{\i}\address{Department of Physics, College of William and Mary, Williamsburg, Virginia 23187 }%
        \thanks{This work was supported in part by the US Department
of Energy under grant No.~DE-FG02-97ER41032. Author thanks F. Gross and J. Tjon
for useful discussions.}}
       
\begin{document}

\maketitle

\begin{abstract}
In nuclear and particle physics one is often faced with problems where 
perturbation theory is not applicable. An example of this is the description 
of bound states. Therefore, an exact solution of field theory to 
all orders is an unavoidable and interesting problem. Path integrals provide 
a framework for exact solutions in field theory. In this talk I will 
present an economical method of evaluating path integrals using the 
Feynman-Schwinger representation (FSR). 
\end{abstract}

\section{Introduction}

The basic idea in the Feynman-Schwinger representation is to replace path 
integrals over quantum fields with path integrals over particle trajectories. 
The step of going from field configurations to particle trajectories 
dramatically reduces the number of degrees of freedom.

In this talk we present analytic and numerical applications of 
Feynman-Schwinger representation~\cite{FEYNMAN,SIMONOV1,SIMONOV2,TJON1,BRAMBILLA,SAVKLI1,SAVKLI2,SAVKLI3} to nonperturbative problems. The first 
applictaion is the calculation of 1-body propagator in massive scalar qed. The
second application involves the calculation of 2-body bound state masses in 
scalar $\chi^2\phi$ interaction.

\section{1-body propagator in massive SQED}

Massive scalar QED in 0+1 dimension is a simple interaction that enables one
to obtain a fully analytical result for the dressed and bound state masses
within the FSR approach. In 0+1 dimension {\em contribution of the matter 
loops identically vanish and therefore quenched FSR 
calculations give the exact result}~\cite{SAVKLI1}. In this section we compare the self 
energy result obtained by three different approaches; namely the simple bubble
 sum, the Dyson-Schwinger equation, and the Feynman-Schwinger representation.

\begin{figure}
\begin{center}
\mbox{
    \epsfxsize=3.2in
\epsffile{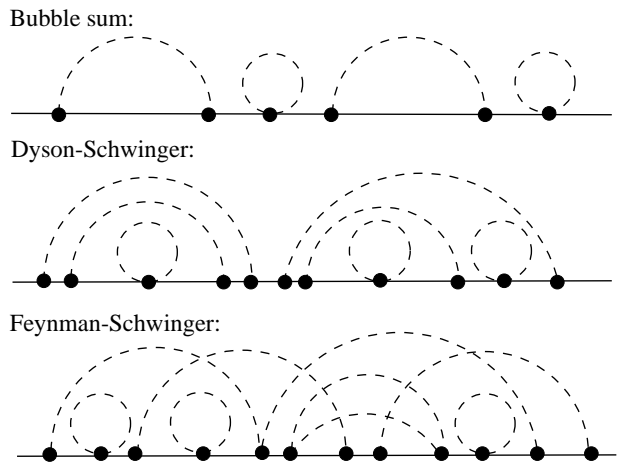}
}
\end{center}
\caption{ Various interactions included in each approach are shown.
The Feynman-Schwinger approach includes all diagrams. In 1-dimension the 
contribution of diagrams with loops of charged particles identically vanishes.~\cite{SAVKLI1} }
\label{allthree.sqed}
\end{figure}

The Minkowski metric expression for the scalar QED Lagrangian in Feynman
gauge is given by
\begin{eqnarray}
{\cal
L}_{SQED}&=&-m^2\chi^2-\frac{1}{4}F^2+\frac{1}{2}\mu^2A^2-\frac{1}{2}( 
\partial A
)^2+(\partial_\mu-ieA_\mu)\chi^*(\partial^\mu+ieA^\mu)\chi,
\end{eqnarray}
where $A$ represents the gauge field of mass $\mu$, and $\chi$ is
the charged field of mass $m$.  The field tensor $F$ is zero in 0+1
dimensions, and the dynamics is described by the gauge fixing term 
$(\partial A)^2$. 
The kind of diagrams included in each method are displayed in
Fig.~\ref{allthree.sqed}. The main difference between the rainbow
Dyson-Schwinger and the Feynman-Schwinger diagrams is the crossed
diagrams. These diagrams involve photon lines that cross each other. The FSR
approach also includes all possible four-point interaction contributions
while the rainbow DSE only includes the tadpole type four-point interactions.
In principle all four-point interactions can also be incorporated into the
simple bubble sum and the rainbow DSE.

\begin{figure}[thb]
\begin{center}
\mbox{
    \epsfxsize=5.0in
\epsffile{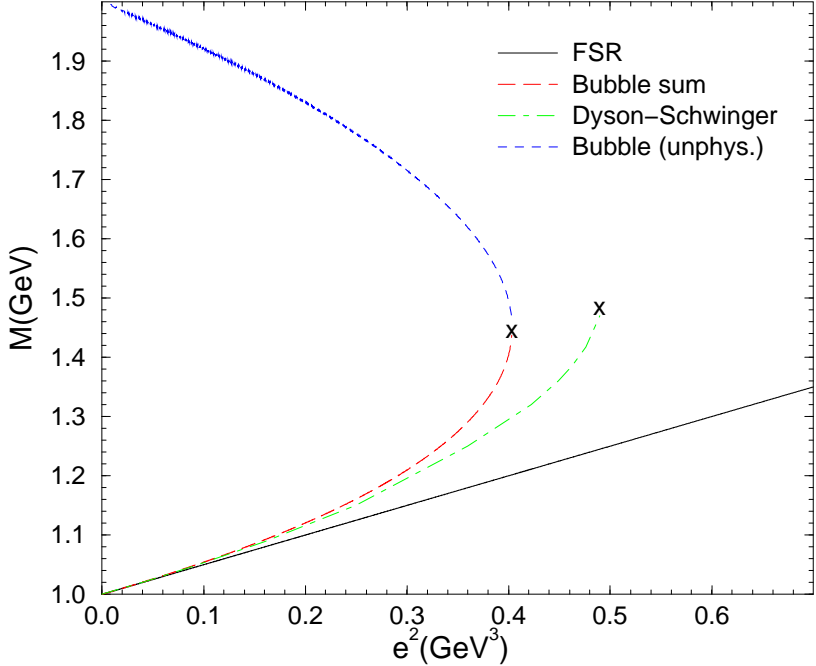}
}
\end{center}
\caption{The 1-body mass calculated by the FSR approach, the
Dyson-Schwinger equation, and the bubble summation for values of $m=\mu=1$ 
GeV. Mass results obtained by bubble summation and the rainbow Dyson-Schwinger 
equations significantly deviate from the exact result provided by the FSR 
method. 
}
\label{mvsg2.sqed}
\end{figure}

In Fig.~\ref{mvsg2.sqed} we display all dressed mass results. The bubble
summation develops a complex mass pole beyond a critical coupling
$e^2_{crit}=0.4$ (GeV)$^3$.
At the critical point a  `collision' takes place with another real solution, 
leading to two complex conjugated solutions with increasing $e^2$. The result 
obtained from the Dyson-Schwinger Equation displays a similar characteristic. At low
coupling strengths the rainbow DS and the bubble results are very  close and 
they converge to the exact result given by the Feynman-Schwinger  approach.
Similar to the bubble result the DS result develops a complex mass 
pole at a critical coupling of $e^2_{crit}=0.49$ (GeV)$^3$. These results 
clearly show that the solution of {\em the Dyson-Schwinger Equation in rainbow
approximation is not a good approximation to the exact result}.
The second application of the FSR involves bound states in scalar $\chi^2\phi$
interaction.
\section{Two-body bound states in scalar $\chi^2\phi$}
The application of the FSR to scalar particles have been considered in References~\cite{TJON1,SAVKLI2,SAVKLI3}. The Euclidean Lagrangian for this theory is given by 
\begin{equation}
{\cal L}_E=\chi^*\bigl[m^2-\partial^2+g\phi\bigr]\chi+\frac{1}{2}\,\phi(\mu^2-\partial^2)\phi.
\label{lagr0}
\end{equation}
Here we present the results for the 2-body bound states.
The final result for the two-body propagator involves a quantum mechanical 
path integral that sums up contributions coming from all possible {\em trajectories} of {\em particles}
\begin{equation}
G=-\int_0^\infty ds \int_0^\infty d\bar{s} \int ({\cal D}z)_{xy}\int ({\cal D}\bar{z})_{\bar{x}\bar{y}}\,e^{-S[Z]},
\label{g1.phi3}
\end{equation}
where $S[Z]$ term involves kinetic and interaction energy contributions. 
Further details on the treatment of scalar interactions can be found in 
Refs.~\cite{SAVKLI2,SAVKLI3}. The interaction kernel $\Delta(x)$ is defined by
\begin{eqnarray}
\Delta(x,\mu)&=& \int \frac{d^4p}{(2\pi)^4}\frac{e^{ip\cdot x}}{p^2+\mu^2}=\frac{\mu}{4\pi^2|x|}K_1(\mu|x|).
\label{kernel.phi3}
\end{eqnarray}
where $\mu$ is chosen to be $\mu=0.15 GeV$.
The ultraviolet singularity in the kernel $\Delta(x,\mu)$ Eq.~(\ref{kernel.phi3}) can be regularized using a Pauli-Villars regularization prescription. The 
Pauli-Villars mass we choose is $M_{PV}=3\mu$.
\begin{figure}[thb]
\begin{center}
\mbox{
   \epsfxsize=5.0in
\epsfbox{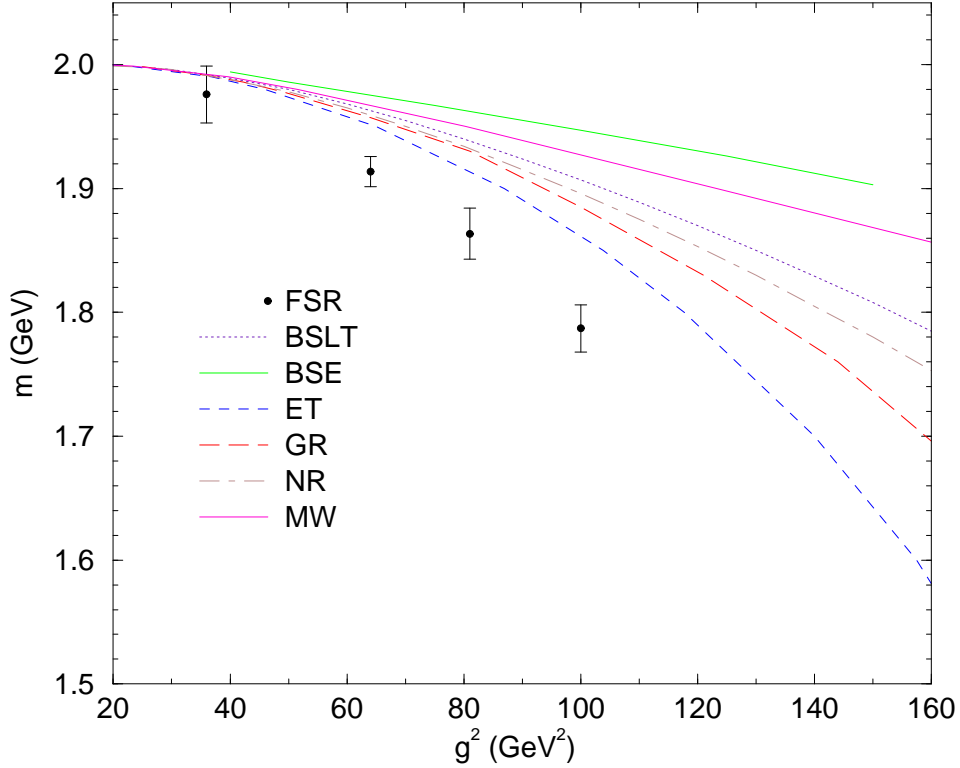}
}
\end{center}
\caption{The coupling constant dependence of the 2-body bound state mass is 
shown. The Bethe-Salpeter equation in ladder approximation gives the poorest 
result (BSE), while the Gross equation (GR) gives the strongest binding among 
manifestly covariant equations. Inclusion of retardation effects push the 
Equal-time result(ET) significantly up (ET with retardation shown as MW). }
\label{mvsg2.2b}
\end{figure}
In Figure~\ref{mvsg2.2b} we present the 
comparison of the 2-body bound state masses obtained by the FSR to various 
bound state equations. The FSR calculation involves summation of all ladder and crossed ladder diagrams, and excludes the self energy contributions.
According to Figure~\ref{mvsg2.2b} all bound state equations underbind. Among the manifestly covariant equations the 
Gross equation gives the closest result to the exact calculation obtained by 
the FSR method. This is due to the fact that in the limit of infinitely 
heavy-light systems the Gross equation effectively sums all ladder and 
crossed ladder diagrams. Equal-time equation also produces a strong binding 
but the inclusion of retardation effects pushes the Equal-time results away 
from the exact results (Mandelzweig-Wallace equation). In particular the Bethe-Salpeter equation in the ladder approximation (BSE in Figure~\ref{mvsg2.2b})
gives the lowest binding. A comparison of the ladder Bethe-Salpeter, Gross, 
and the FSR results shows that {\em the exchange of crossed ladder diagrams 
plays a very significant role.}
 
\section{Conclusion}
Results presented in this talk clearly shows that {\em approximate calculations
in field theory may lead to serious deviations from the exact results.}
Therefore it is important to develop rigorous nonperturbative methods. The 
FSR is a promising candidate for doing nonperturbative calculations in field 
theory.


\begin{thebibliography}{9}
\bibitem{FEYNMAN} R.P. Feynman, Phys. Rev 80 (1950), 440; J. Schwinger, Phys. Rev. {\bf 82} (1951), 664
\bibitem{SIMONOV1} Yu. A. Simonov, Nucl. Phys. B {\bf 307} (1988), 
\bibitem{SIMONOV2} Yu. A. Simonov and J. A. Tjon, Ann. Phys. {\bf 228}, 1 (1993).
\bibitem{TJON1} T. Nieuwenhuis, J. A. Tjon, Phys. Rev. Lett. {\bf 77}, 814 (1996)
\bibitem{BRAMBILLA} N. Brambilla, A. Vairo, Phys. Rev D. {\bf 56}, 1445 (1997).
\bibitem{SAVKLI1}\c{C} \c{S}avkl{\i}, F. Gross, J. A. Tjon, hep-ph/9907445, accepted for publication in Phys. Rev. D.
\bibitem{SAVKLI2} \c{C} \c{S}avkl{\i}, hep-ph/9910502, accepted for 
publication in Comp. Phys. Comm.
\bibitem{SAVKLI3} \c{C}. \c{S}avkl{\i}, J. A. Tjon, F. Gross, Phys. Rev. C 
{\bf 60} : 055210, 1999, Erratum-ibid. {\bf C} 61 : 069901, 2000. 
\end{thebibliography}
\end{document}